\documentclass[12pt]{article}
\usepackage[cp1251]{inputenc}
\usepackage[english]{babel}
\usepackage{amsfonts,amssymb}
\usepackage{graphicx}
\usepackage{amsmath, amssymb, graphics, setspace}
\usepackage{hyperref}

\begin{document}

\title{The mass function method for obtaining exact solutions of Einstein equations.}
\maketitle
\begin{center}
M.~Korkina$^1$ \quad E.~Kopteva$^1$ \\
       $^1${\it Theoretical Physics Department, Dnepropetrovsk National University,
Dniepropetrovsk, Ukraine} \\
\end{center}

\begin{abstract}
We show that the mass function method makes it much easier to obtain the exact solutions of Einstein equations. The known solutions for empty space and for the universe filled with dust-like matter are considered from the point of view of the method and new exact solutions are obtained. Generalized solutions for the Schwarzschild type black holes are considered. It is shown that such black holes always must contain non-baryonic matter. New exact solution for the black hole embedded into dust matter cosmological background is obtained.
\end{abstract}

\subsection*{1. Introduction}

In spite of the fact that the area of investigations in general relativity concerning the finding of exact solutions is broadly developed \cite{1}, this research direction stays still actual. As it was noticed yet by  J. Synge: «In a complicated situation such as we have before us, an exact solution of the field equations is to be esteemed far above any approximation, even though the exact mathematical solution is admittedly only an approximation to physical reality (no mathematical formula is ever more than this anyway)» \cite{2}. In this paper we consider the mass function method for obtaining exact spherically symmetric solutions of Einstein equations.

The mass function is one of four algebraic invariants existing for spherically symmetric metric \cite{3}. For the interval of general form
\begin{equation}\label{GrindEQ__1_} 
ds^{2} =e^{\nu (R,t)} dt^{2} -e^{\lambda (R,t)} dR^{2} -r^{2} (R,t)d\sigma ^{2}
\end{equation}
the mass function is defined as follows
\begin{equation}\label{GrindEQ__2_} 
m(R,t)=r(R,t)\left(1+e^{-\nu (R,t)} \dot{r}^{2} -e^{-\lambda (R,t)} r'^{2} \right),
\end{equation}
where prime means partial derivative with respect to coordinate $R$, and dot means partial derivative with respect to time $t$; here and further $c=1$, $d\sigma ^{2} $ is a standard metric on 2-sphere. 
The mass function was previously mentioned in papers \cite{4} -- \cite{7} and also known as Misner-Sharp mass.

By use of mass function it is possible to rewrite the Einstein equations in much more simpler way:
\begin{equation}\label{GrindEQ__3_}
m'=\varepsilon r^{2} r';
\end{equation}
\begin{equation}\label{GrindEQ__4_}
\dot{m}=-p_{||} r^{2} \dot{r};
\end{equation}
\begin{equation}\label{GrindEQ__5_}
2\dot{r}'=\nu '\dot{r}+\dot{\lambda }r';
\end{equation}
\begin{equation}\label{GrindEQ__6_}
2\dot{m}'=m'\frac{\dot{r}}{r'} \nu '+\dot{m}\frac{r'}{\dot{r}} \dot{\lambda }
-4r\dot{r}r'p_{\bot } ,
\end{equation}
where $\varepsilon $ means the energy density including factor ${8\pi \gamma  \mathord{\left/{\vphantom{8\pi \gamma  c^{4} }}\right.\kern-\nulldelimiterspace} c^{4} } $ so that it has a dimension of  $cm^{-2} $, $p_{||} $ is radial component of pressure and $p_{\bot } $ is tangential one ($T_{2}^{2} =-p_{\bot } $) in the same units  $cm^{-2} $. In the system (\ref{GrindEQ__3_} -- \ref{GrindEQ__6_}) the equation (\ref{GrindEQ__3_}) corresponds to the equation for   $T_{0}^{0} $: $R_{0}^{0} -{R \mathord{\left/{\vphantom{R 2}}\right.\kern-\nulldelimiterspace} 2} =\varepsilon $, the equation (\ref{GrindEQ__4_}) corresponds to the one for  $T_{1}^{1} $, the equation (\ref{GrindEQ__5_}), is in fact a condition of comoving frame $T_{0}^{1} =0$ , the equation (\ref{GrindEQ__6_}) follows from the equation for $T_{2}^{2} $.

\subsection*{2. Spherically symmetric solutions for the empty space}

In this section we shall demonstrate the mass function method application for obtaining the spherically symmetric solutions for the empty space. In this case from (\ref{GrindEQ__3_}) and (\ref{GrindEQ__4_}) it follows that  $m(R,t)=const=r_{g} $, and the equation (\ref{GrindEQ__6_}) becomes identically null. Thus the complete system of equations for the empty space takes the form
\begin{equation}
\label{GrindEQ__7_} r_{g} =r(R,t)\left(1+e^{-\nu (R,t)} \dot{r}^{2} -e^{-\lambda (R,t)} r'^{2} \right);
\end{equation}
\begin{equation}
\label{GrindEQ__8_} 2\dot{r}'=\nu '\dot{r}+\dot{\lambda }r'.
\end{equation}
One should stress here that for the case under consideration the equations (\ref{GrindEQ__7_}) and (\ref{GrindEQ__8_}) completely exhaust all the system of Einstein equations. In order to obtain the solution of the system (\ref{GrindEQ__7_}), (\ref{GrindEQ__8_}), one should choose the coordinate conditions.

Let us first take the synchronous coordinate system $e^{\nu (R,t)} =1$.

Under such coordinate condition it follows from (\ref{GrindEQ__8_}) that the expression $e^{-\lambda (R,t)} r'^{2} $ depends on coordinate $R$ only: $e^{-\lambda } r'^{2} \equiv f^{2} (R)$. Hence one obtains from (\ref{GrindEQ__7_}) the equation
\begin{equation}
\label{GrindEQ__9_} r_{g} =r(R,t)\left(1+\dot{r}^{2} -f^{2} (R)\right).
\end{equation}
This equation can be easily integrated. The three resulting solutions ($f^{2} (R)>,=,<1$) will be particular case of the well known Tolman-Bondi solution \cite{8}, that will be discussed further.

Let us now take Gaussian coordinate system $e^{\lambda (R,t)} =1$.

Under such coordinate condition we obtain the equation analogous to (\ref{GrindEQ__9_}):

\begin{equation}
\label{GrindEQ__10_} r_{g} =r(R,t)\left(1+\psi ^{2} (t)-r'^{2} \right),
\end{equation}
where $\psi ^{2} (t)$ is an arbitrary function of integration. The solution of the equation (\ref{GrindEQ__10_}) has the form:

\begin{equation} \label{GrindEQ__11_} 
ds^{2} =\tanh ^{2} \frac{\alpha }{2} dt^{2} -dR^{2} -r_{g} ^{2} \cosh ^{4} \frac{\alpha }{2} d\sigma ^{2} , 
\end{equation}
where

\[R=\frac{r_{g} }{2} \left(\sinh \alpha +\alpha \right),\]
with $\alpha $ being a dimensionless parameter.

If one chooses the coordinate conditions so that the spatial part of the metric will be conformally flat, i.e.

\begin{equation} \label{GrindEQ__12_} ds^{2} =e^{\nu (R,t)} dt^{2} -e^{\lambda (R,t)} \left(dR^{2} +R^{2} d\sigma ^{2} \right), \end{equation}
then one will obtain

\begin{equation} \label{GrindEQ__13_} r(R,t)=e^{\frac{\lambda }{2} } R. \end{equation}
From where one has

\[\dot{\lambda }=\frac{2\dot{r}}{r} ;\]
and
\[2\dot{r}'=\nu '\dot{r}+\frac{2\dot{r}}{r} r';\]

\begin{equation} \label{GrindEQ__14_} e^{\nu } =\frac{\dot{r}^{2} }{r^{2} \psi ^{2} (t)} . \end{equation}

Grouping (\ref{GrindEQ__7_}) with (\ref{GrindEQ__14_}) one has

\[r_{g} =r\left(1+r^{2} \psi ^{2} (t)-\frac{R^{2} }{r^{2} } r'^{2} \right),\]
or consequently 
\begin{equation} \label{GrindEQ__15_} \frac{dr}{\sqrt{r^{4} \psi ^{2} (t)+r^{2} -rr_{g} } } =\frac{dR}{R} . \end{equation}

In general case this equation cannot be integrated in elementary functions, but in particular case when $r=r(R)$ it is possible to rewrite:

\[r_{g} =r\left(1-e^{-\lambda } r'^{2} \right).\]
Taking into account (\ref{GrindEQ__13_}) finally we obtain

\begin{equation} \label{GrindEQ__16_} 
r(R)=\left(R+\frac{r_{g} }{4R} \right)^{2} . 
\end{equation}

Another result comes from the system of equations (\ref{GrindEQ__7_}), (\ref{GrindEQ__8_}) under coordinate conditions in the form:

\begin{equation} \label{GrindEQ__17_} 
\begin{array}{c} {e^{\nu (R,t)} =e^{\lambda (R,t)} {\rm \; ,}} \\ {\lambda (R,t)=\lambda (r(R,t)){\rm \; .}} \end{array} 
\end{equation}
This is the generalization of the known Kruskal-Szekeres solution \cite{1}. From (\ref{GrindEQ__8_}) with (\ref{GrindEQ__17_}) the following equation arises 

\begin{equation} \label{GrindEQ__18_} 
\dot{r}'=\dot{r}r'\frac{d\lambda }{dr} . 
\end{equation}
After integration of (\ref{GrindEQ__18_}) with respect to $t$ and $R$ we obtain

\begin{equation} \label{GrindEQ__19_} \begin{array}{c} {\dot{r}(R,t)=e^{\lambda } \psi (t){\rm \; ;}} \\ {r'(R,t)=e^{\lambda } K(R){\rm \; ,}} \end{array} 
\end{equation}
with arbitrary dimensionless functions of integration $\psi (t)$ and $K(R)$. Then we input the expressions from (\ref{GrindEQ__19_}) into the equation (\ref{GrindEQ__7_}):

\begin{equation} \label{GrindEQ__20_} 
\left(\frac{r_{g} }{r} -1\right)e^{-\lambda (r)} =\psi ^{2} (t)-K^{2} (R), 
\end{equation}

From (\ref{GrindEQ__20_}) it follows that $r(R,t)=r(\alpha )$, where $\alpha \equiv \psi ^{2} (t)-K^{2} (R)$.

Let us consider the particular case of the solution (\ref{GrindEQ__20_}). If we take the arbitrary functions in the form

\begin{equation} \label{GrindEQ__21_} \begin{array}{l} {\psi ^{2} (t)=\frac{t^{2} }{4r_{g}^{2} } {\rm \; ;}} \\ {K^{2} (R)=\frac{R^{2} }{4r_{g}^{2} } {\rm \; ,}} \end{array} \end{equation}
then from (\ref{GrindEQ__20_}) we will obtain the known Kruskal-Szekeres solution:

\begin{equation} \label{GrindEQ__22_} \left(\frac{r_{g} }{r} -1\right)e^{-\lambda (r)} =\frac{t^{2} -R^{2} }{4r_{g}^{2} } =\alpha . \end{equation}
Inputting (\ref{GrindEQ__21_}) into (\ref{GrindEQ__19_}) we find the expression for $e^{\lambda } $ and substitute it into (\ref{GrindEQ__22_}). Then after separation of variables and integration we will have the result

\begin{equation} \label{GrindEQ__23_} C\alpha =e^{\frac{r}{r_{g} } } \left(r_{g} -r\right). \end{equation}

Thus from (\ref{GrindEQ__22_}) and (\ref{GrindEQ__23_}) we have the expression for $e^{\lambda } $:

\begin{equation} \label{GrindEQ__26_} e^{\lambda } =\frac{C}{r} e^{-{r \mathord{\left/{\vphantom{r r_{g} }}\right.\kern-\nulldelimiterspace} r_{g} } } . \end{equation}
And finally we obtain the required metric in the form

\begin{equation} \label{GrindEQ__27_} ds^{2} =\frac{C}{r} e^{-{r \mathord{\left/{\vphantom{r r_{g} }}\right.\kern-\nulldelimiterspace} r_{g} } } \left(dt^{2} -dR^{2} \right)-r^{2} d\sigma ^{2} , \end{equation}
where
\[e^{{r \mathord{\left/{\vphantom{r r_{g} }}\right.\kern-\nulldelimiterspace} r_{g} } } \left(r-r_{g} \right)=\frac{C}{4r_{g}^{2} } \left(t^{2} -R^{2} \right).\]

\subsection*{3. Solutions for the dust matter}

The dust matter is characterized by zero pressure $p=0$. In this case the equation (\ref{GrindEQ__4_}) gives $\dot{m}=0$, that means that the mass function depends on spatial coordinate only. Thus from (\ref{GrindEQ__6_}) it follows that $\nu '=0$, and hence the coordinate system becomes synchronous. The equation (\ref{GrindEQ__5_}) then yields the expression $e^{-\lambda } r'^{2} \equiv f^{2} (R)$, where $f(R)$ is arbitrary function of integration. The mass function (\ref{GrindEQ__2_}) in this case will have the form:

\begin{equation} \label{GrindEQ__28_} 
m(R)=r(R,t)\left(1+\dot{r}^{2} (R,t)-f^{2} (R)\right). 
\end{equation}

The equation (\ref{GrindEQ__28_}) immediately leads to the Tolman-Bondi solution. Indeed, expressing  $\dot{r}$ from (\ref{GrindEQ__28_}) and integrating in standard way one obtains three types of the Tolman-Bondi solution depending on the sign of $f^{2} (R)-1$ for the interval in new terms

\begin{equation} \label{GrindEQ__29_} 
ds^{2} =dt^{2} -\frac{r'^{2} (R,t)}{f^{2} (R)} dR^{2} -r^{2} (R,t)d\sigma ^{2}. 
\end{equation}

The hyperbolic type of solution ($f^{2} (R)>1$) reads

\begin{equation} \label{GrindEQ__30_} \begin{array}{c} {r(R,t)=\frac{m(R)}{f^{2} (R)-1} \sinh ^{2} \frac{\alpha }{2} ;} \\ {t-t_{0} (R)=\pm \frac{m(R)}{2(f^{2} (R)-1)^{{3 \mathord{\left/{\vphantom{3 2}}\right.\kern-\nulldelimiterspace} 2} } } (\sinh \alpha -\alpha );} \end{array} \end{equation}

The elliptic type of solution ($f^{2} (R)<1$) reads

\begin{equation} \label{GrindEQ__31_} \begin{array}{c} {r(R,t)=\frac{m(R)}{1-f^{2} (R)} \sin ^{2} \frac{\alpha }{2} ;} \\ {t-t_{0} (R)=\frac{m(R)}{2(1-f^{2} (R))^{{3 \mathord{\left/{\vphantom{3 2}}\right.\kern-\nulldelimiterspace} 2} } } (\alpha -\sin \alpha );} \end{array} 
\end{equation}

The parabolic type of solution ($f^{2} (R)=1$) reads:

\begin{equation} \label{GrindEQ__32_} r(R,t)=\left[\pm \frac{3}{2} \sqrt{m(R)} (t-t_{0} (R))\right]^{\frac{2}{3} } . 
\end{equation}

Here the mass function $m(R)$ is arbitrary function of $R$, and has the sense of total mass of the dust distribution within the shell $R=const$.

As known the Friedmann solution for the dust matter is the particular case of the Tolman-Bondi solution under certain choice of arbitrary functions $m(R)$, $f(R)$ and $t_{0} (R)$, so it can be easily obtained by the mass function method. Moreover, significant simplification takes place while applying the method for obtaining the Friedman type solutions for the case of nonzero pressure that depends on time only. Let us consider this case in more details. 

For the homogeneous and isotropic interval

\begin{equation} \label{GrindEQ__33_} ds^{2} =dt^{2} -a^{2} (t)\left(dR^{2} +F^{2} (R)d\sigma ^{2} \right), 
\end{equation}
where $F^{2} (R)=\sin ^{2} R,{\rm \; \; }\sinh ^{2} R,{\rm \; \; }R^{2} $ for the closed, open and flat space, respectively, the mass function concerning (\ref{GrindEQ__1_}) and (\ref{GrindEQ__2_}) has the form

\begin{equation} \label{GrindEQ__34_} m(t,R)=a(t)F^{3} (R)\left(\dot{a}^{2} (t)+k\right), \end{equation}
where $k=0,{\rm \; }\pm 1$ is curvature parameter. Rewriting the system of equations (\ref{GrindEQ__3_} -- \ref{GrindEQ__6_}) for the interval (\ref{GrindEQ__33_}) with mass function (\ref{GrindEQ__34_}) we obtain the Friedmann equations for the matter with pressure:

\begin{equation} \label{GrindEQ__35_} 
\frac{1}{a^{2} (t)} \left(\dot{a}^{2} (t)+k\right)=\frac{8\pi \gamma }{c^{4} } \frac{\varepsilon (t)}{3} ; 
\end{equation}

\begin{equation} \label{GrindEQ__36_} 
\frac{\dot{\varepsilon }(t)}{\varepsilon (t)+p(t)} =-\frac{3\dot{a}(t)}{a(t)} . 
\end{equation}

In general case for arbitrary dependence  $p(t)$ it is impossible to find the exact solution of the system (\ref{GrindEQ__35_}), (\ref{GrindEQ__36_}). But a set of important physically meaningful solutions can be found under supposition that 

\begin{equation} \label{GrindEQ__37_} 
p(t)=n\varepsilon (t), 
\end{equation}
where $n$ is arbitrary constant.

After substitution (\ref{GrindEQ__36_}) into (\ref{GrindEQ__37_}) and integrating we obtain:

\begin{equation} \label{GrindEQ__38_} \varepsilon (t)=\frac{1}{a_{n}^{2} } \frac{3}{\left[{a(t) \mathord{\left/{\vphantom{a(t) a_{n} }}\right.\kern-\nulldelimiterspace} a_{n} } \right]^{3(n+1)} } , 
\end{equation}
where $a_{n} $ -- is constant of integration.
Inputting the energy density (\ref{GrindEQ__38_}) into the equation (\ref{GrindEQ__35_}) and separating variables we have:

\begin{equation} \label{GrindEQ__39_} \frac{da}{\sqrt{\left[{a_{n}  \mathord{\left/{\vphantom{a_{n}  a}}\right.\kern-\nulldelimiterspace} a} \right]^{3n+1} -k} } =dt. 
\end{equation}
For the Lobachevsky space ($k=-1$) the solution of the equation (\ref{GrindEQ__39_}) reads

\begin{equation} \label{GrindEQ__391_} ds^{2} =a_{n}^{2} \sinh ^{2\beta _{n} } \frac{\alpha }{\beta _{n} } \left(d\alpha ^{2} -dR^{2} -\sinh ^{2} Rd\sigma ^{2} \right),
\end{equation}
with $\alpha $ being a parameter:

\[dt=a(\alpha )d\alpha ,\]

\[a(\alpha )=a_{n} \sinh ^{\beta _{n} } \frac{\alpha }{\beta _{n} } ,\]
and the constant of integration is chosen as $\beta _{n} =\frac{2}{3n+1} $.

Analogous solution for the closed world has the form

\begin{equation} \label{GrindEQ__392_} ds^{2} =a_{n}^{2} \sin ^{2\beta _{n} } \frac{\alpha }{\beta _{n} } \left(d\alpha ^{2} -dR^{2} -\sin ^{2} Rd\sigma ^{2} \right),
\end{equation}
and for the flat space one has

\begin{equation} \label{GrindEQ__393_} ds^{2} =a_{n}^{2} \left(\frac{1+\beta _{n} }{\beta _{n} } \cdot t\right)^{\frac{2\beta _{n} }{1+\beta _{n} } } \left(d\alpha ^{2} -dR^{2} -R^{2} d\sigma ^{2} \right).
\end{equation}

\subsection*{4. Exact solutions for Schwarzschild-like black holes}

Let us now consider the class of static spherically-symmetric solutions in curvature coordinates \cite{Gautreau}:

\begin{equation} \label{GrindEQ__40_} 
ds^{2} =e^{\nu (r)} dt^{2} -e^{\lambda (r)} dR^{2} -r^{2} d\sigma ^{2} . 
\end{equation}

For the metric (\ref{GrindEQ__40_}) the mass function reads

\begin{equation} \label{GrindEQ__41_} 
m(r)=r\left(1-e^{-\lambda (r)} \right). 
\end{equation}

And the energy density in this case according to (\ref{GrindEQ__3_}) will be

\begin{equation} \label{GrindEQ__42_} 
\varepsilon (r)=\frac{m'(r)}{r^{2} } . 
\end{equation}

Let us take anisotropic fluid as a source. Thus we have the stress-energy tensor with following nonzero components: $T_{0}^{0} =\varepsilon $, $T_{1}^{1} =-p_{||} $, $T_{2}^{2} =T_{3}^{3} =-p_{\bot } $. In order to obtain an exact solution for the case under consideration we will fix the coordinate conditions as follows:

\begin{equation} \label{GrindEQ__43_} 
e^{\nu (r)} =e^{-\lambda (r)} . 
\end{equation}

Taking into account  (\ref{GrindEQ__43_}) and (\ref{GrindEQ__41_}) one can rewrite the metric  (\ref{GrindEQ__40_}) 

\begin{equation} \label{GrindEQ__44_} 
ds^{2} =\left(1-\frac{m(r)}{r} \right)dt^{2} -\left(1-\frac{m(r)}{r} \right)^{-1} dr^{2} -r^{2} d\sigma ^{2} . 
\end{equation}

Einstein equations for the components  $T_{0}^{0} $  and  $T_{1}^{1} $  under condition  (\ref{GrindEQ__43_}) will lead to the equality 

\begin{equation} \label{GrindEQ__45_} 
T_{0}^{0} =T_{1}^{1} , 
\end{equation}
that means that the radial pressure is always negative  $p_{||} =-T_{1}^{1} $. Thus one can conclude that under condition (\ref{GrindEQ__43_}) the space-time (\ref{GrindEQ__44_}) always contains non-baryonic matter.

Let us further consider the stress-energy conservation law $T_{\nu ;\mu }^{\mu } =0$ which is more suitable here than the rest equation for  $T_{2}^{2} $. For our case this gives  

\begin{equation} \label{GrindEQ__46_} 
\frac{\partial }{\partial r} T_{1}^{1} +\frac{2}{r} T_{1}^{1} +\frac{2}{r} T_{2}^{2} =0, 
\end{equation}
here we take into account that  $T_{2}^{2} =T_{3}^{3} $ for the spherical symmetry. Substituting radial pressure expressed from (\ref{GrindEQ__42_}) and (\ref{GrindEQ__45_}) into (\ref{GrindEQ__46_}) one has for the tangent pressure:

\begin{equation} \label{GrindEQ__47_} 
T_{2}^{2} =\frac{m''(r)}{2r} =-p_{\bot } . 
\end{equation}

So all the metric coefficients and nonzero components of the stress-energy tensor are expressed in terms of the mass function and its derivatives. The state equation in this case  (under condition (\ref{GrindEQ__45_})) will be the relation between radial and tangent parts of pressure, and it will completely define the solution.

The metric (\ref{GrindEQ__41_}) allows different types of solutions concerning their physical properties.

1.	Under $m(r)<r$  for all $r\in(0<r<\infty) $ these solutions describe the only R-region. As far as radial pressure is negative these solutions can possibly describe the spherical configurations with «dark energy».

2.	Under  $m(r)>r$ for all $r\in(0<r<\infty) $  the solutions describe the only T-region. For example the state equation $T_{2}^{2} =\frac{r}{2} T_{1}^{1} $  leads to the following T-solution: 

\begin{equation} \label{GrindEQ__48_} 
ds^{2} =\left(\frac{r_{0} }{t} e^{\frac{t}{r_{0} } } -1\right)^{-1} dt^{2} -\left(\frac{r_{0} }{t} e^{\frac{t}{r_{0} } } -1\right)dx^{2} -t^{2} d\sigma ^{2} , 
\end{equation}
where $r_{0} $ is an arbitrary constant.

3.	If the metric coefficients $e^{\nu (r)} $  and $e^{\lambda (r)} $  are alternating functions the solutions will describe the Schwarzschild-like black holes. Let us suppose the equality  

\begin{equation} \label{GrindEQ__49_} T_{2}^{2} =\beta T_{1}^{1} , 
\end{equation}
with $\beta $ being arbitrary constant. From (\ref{GrindEQ__42_}) and (\ref{GrindEQ__47_}) under (\ref{GrindEQ__49_}) one obtains the following equation for the mass function

\begin{equation} \label{GrindEQ__50_}
m''(r)-2\beta m'(r)\frac{1}{r} =0, 
\end{equation}
which has the solution

\[m(r)=C_{1} \frac{r^{2\beta +1} }{2\beta +1} +C_{2} .\]

Let us choose the constants to be  $C_{2} =r_{g} $ ($r_{g} =\frac{2\gamma M}{c^{2} } $ is Schwarzschild radius), $\frac{C_{1} }{2\beta +1} =A$ ($2\beta +1\ne 0$), and rewrite the mass function

\begin{equation} \label{GrindEQ__51_} 
m(r)=r_{g} +Ar^{2\beta +1} . 
\end{equation}

 $\beta =1$  in (\ref{GrindEQ__51_}) gives either de Sitter solution \cite{9} under  $r_{g} =0$ or the Kottler solution \cite{10} under nonzero  $r_{g} $. If $\beta =-1$  there will be Reissner-Nordstrem solution \cite{11}. If $\beta =\frac{1}{2} $ one will have the solution obtained in  \cite{12, 13}, which describes the Rindler space-time which is actively discussed last time concerning the interior of black holes  \cite{14, 15}. The case of  $\beta =-\frac{1}{2} $ is special, it gives   $e^{-\lambda } =1-\frac{r_{g} }{r} -\frac{r_{0} }{r} \ln \frac{r}{r_{0} } $.

For all $\beta \ne -\frac{1}{2} $ it is possible to write the solution in general form

\begin{equation} \label{GrindEQ__52_} 
ds^{2} =\left(1-\frac{r_{g} }{r} -\sum _{i}A_{i} r^{2\beta _{i} }  \right)dt^{2} -\left(1-\frac{r_{g} }{r} -\sum _{i}A_{i} r^{2\beta _{i} }  \right)^{-1} dr^{2} -r^{2} d\sigma ^{2} . 
\end{equation}
Here we take into account the additivity of mass function so that if one has several sources their contributions to the metric coefficients in (\ref{GrindEQ__52_}) will be summarised  (under supposition of weak interaction between the sources  $\varepsilon =\sum _{i}\varepsilon _{i}  $, $p=\sum _{i}p_{i}  $).

The physical sense of different $\beta $  is not obvious. But in  \cite{16} the black hole was considered surrounded by quintessence that corresponds to  $\beta =-\frac{1}{3} $. The problem of investigation of the general properties of such black holes is a question of our further great interest.

\subsection*{5.	The solutions for the black holes on the dust matter background}

The problem of building a model of the black hole that embedded into space which is not empty but filled with some matter is of great interest in wide set of research directions, including the thermodynamics of black holes \cite{Giddings},\cite{FirouzjaeeMansouri},\cite{FirouzjaeeEllis},\cite{FirouzjaeeMansouri2}, the black hole horizon dynamics \cite{FirouzjaeeMansouri3}, studying the influence of cosmological expansion on the evolution of local objects \cite{MoradiFirouzjaeeMansouri},\cite{FaraoniJacques} etc.

One of the pioneer works in this direction is the paper by McVittie  \cite{17}, were the solution was obtained in the fallowing form

\begin{equation} \label{GrindEQ__53_} 
ds^{2} =\left[\frac{1-{r_{g} \mu (t) \mathord{\left/{\vphantom{r_{g} \mu (t) 4R}}\right.\kern-\nulldelimiterspace} 4R} }{1+{r_{g} \mu (t) \mathord{\left/{\vphantom{r_{g} \mu (t) 4R}}\right.\kern-\nulldelimiterspace} 4R} } \right]^{2} dt^{2} -\frac{1}{\mu ^{2} (t)} \left[1+\frac{r_{g} \mu (t)}{4R} \right]^{4} \left(dR^{2} +R^{2} d\sigma ^{2} \right). 
\end{equation}

If $\mu (t)=const=1$ the expression (\ref{GrindEQ__53_}) will give the Schwarzschild metric in isotropic coordinates.  Under  $r_{g} =0$ the metric (\ref{GrindEQ__53_}) takes the form

\begin{equation} \label{GrindEQ__54_} 
ds^{2} =dt^{2} -\frac{1}{\mu ^{2} (t)} \left(dR^{2} +R^{2} d\sigma ^{2} \right), 
\end{equation}
that can be treated  as a Friedman model for the flat space case. Depending on the choice of  $\mu (t)$ it can be either dust model or the model with nonzero pressure. McVittie proposed his solution pretending to be a model of the massive particle in the expanding universe. But in papers \cite{18, 19} it was shown that this solution is unsuitable for the point mass in the universe, although it may possibly be used to describe the black hole.
To study the physical sense of the solution \eqref{GrindEQ__53_} we will find the expression for the correspondent mass function:

\begin{equation} \label{GrindEQ__55_} 
m(R,t)=r_{g} +R^{3} \frac{\dot{\mu }^{2} (t)}{\mu ^{5} (t)} \left(1+\frac{r_{g} }{4R} \mu (t)\right)^{6} . 
\end{equation}

From \eqref{GrindEQ__55_} one has $m=r_{g} $ under $\dot{\mu }(t)=0$, i.e. the mass function for the Schwarzschild solution. And under condition $r_{g} =0$ and $\frac{\dot{\mu }^{2} (t)}{\mu ^{5} (t)} =const\equiv \frac{1}{a_{0}^{2} } $ one has the mass function for the Friedmann solution for the dust. Let us find the energy density and pressure for the metric \eqref{GrindEQ__1_} using \eqref{GrindEQ__3_} and \eqref{GrindEQ__4_}. 

\begin{equation} \label{GrindEQ__56_} 
\begin{array}{c} {\varepsilon (t)=3\frac{\dot{\mu }^{2} (t)}{\mu ^{2} (t)} ;} \\ {p(R,t)=6\mu ^{3} (t)\frac{{r_{g}  \mathord{\left/{\vphantom{r_{g}  4R}}\right.\kern-\nulldelimiterspace} 4R} }{\left({r_{g}  \mathord{\left/{\vphantom{r_{g}  4R}}\right.\kern-\nulldelimiterspace} 4R} \right)-1} .} \end{array} 
\end{equation}

From the expressions \eqref{GrindEQ__55_} and \eqref{GrindEQ__56_} it is clear that the solution \eqref{GrindEQ__53_} can describe neither point mass nor the black hole in the Friedmann space-time (as far as pressure depends on both $R$ and $t$).

The solutions that were obtained in papers \cite{20} -- \cite{22} are in fact the modifications of the solution \eqref{GrindEQ__53_}. So the solution from \cite{20} is \eqref{GrindEQ__53_} with metric coefficients of terms with $dR^{2} $ and $d\sigma ^{2} $ multiplied by some function of time. The solution from \cite{21} and \cite{22}  is \eqref{GrindEQ__53_} with $\mu (t)=const$ and metric coefficients of $dR^{2} $ and $d\sigma ^{2} $ also multiplied by function of time. Analogous consideration of these solutions using the mass function method shows that they also describe neither point mass nor the black hole in the Friedmann or even in Tolman-Bondi space-time because the pressure calculated for them depends both on $R$ and $t$. 

Using the mass function method we will now build the model for the black hole embedded into the Tolman-Bondi universe.

Tolman-Bondi metric for the spherically-symmetric dust distribution in comoving frame (which is synchronous for the dust) has the form (\ref{GrindEQ__29_}).

The Schwarzschild solution is a particular case of the Tolman-Bondi solution under $m(R)=r_{g}$. The Friedmann solution comes from the Tolman-Bondi solution under certain choice of arbitrary functions, for example for the flat case one has $m(R)=a_{0} R^{3} $, $f(R)=1$, $t_{0} (R)=0$.

Thereby the solution for the Schwarzschild-like black hole on the dust cosmological background will be the Tolman-Bondi solution with mass function combined as follows:

\begin{equation} \label{GrindEQ__57_} 
m(R)\to r_{g} +m(R). 
\end{equation}

For the flat case one should take in (\ref{GrindEQ__32_}) the mass function in the form $m(R)= r_{g}+a_{0} R^{3} $ and the solution for the dust universe with embedded Schwarzschild black hole will be

\begin{equation} \label{GrindEQ__58_} 
r(R,t)=\left[\pm \frac{3}{2} \sqrt{r_{g} +a_{0} R^{3} } (t-t_{0} (R))\right]^{\frac{2}{3} } . 
\end{equation}

One should notice that \eqref{GrindEQ__58_} does not describe the pure Friedmann world. This is Tolman-Bondi world with the same mass function as in Friedmann solution. The reason is that $t_{0} (R)$ can not be zero for the flat Schwarzschild solution (unlike $t_{0} (R)=0$ for Friedmann).

The solution obtained in \cite{23}

\begin{equation} \label{GrindEQ__59_} 
r(R,t)=\left[\frac{3}{2} \left(\sqrt{a_{0} R^{3} } +\sqrt{r_{g} } \right)t+R^{\frac{3}{2} } \right]^{\frac{2}{3} }  
\end{equation}
has in round brackets the sum of two correspondent expressions for the Friedmann and Schwarzschild cases with chosen $t_{0} (R)=-R^{{3 \mathord{\left/{\vphantom{3 2}}\right.\kern-\nulldelimiterspace} 2} } a_{0} $. The mass function for this solution reads

\begin{equation} \label{GrindEQ__60_} 
m(R)=a_{0} R^{3} +r_{g} +2\sqrt{a_{0} r_{g} } R^{{3 \mathord{\left/{\vphantom{3 2}}\right.\kern-\nulldelimiterspace} 2} } . 
\end{equation}

One has here the black hole in some specific space-time with mass function \eqref{GrindEQ__60_} but not in the universe filled with dust.

Supposing in \eqref{GrindEQ__31_} $f(R)=\cos R$ as for closed Friedmann solution and taking mass function as $m(R)=r_{g} +a_{0} \sin ^{3} R$ one has

\begin{equation} \label{GrindEQ__61_} 
\begin{array}{c} {r(R,t)=\frac{r_{g} +a_{0} \sin ^{3} R}{\sin ^{2} R} \sin ^{2} \frac{\alpha }{2} ;} \\ {t-t_{0} (R)=\frac{r_{g} +a_{0} \sin ^{3} R}{2\sin ^{3} R} (\alpha -\sin \alpha ).} \end{array} 
\end{equation}

Similarly for \eqref{GrindEQ__30_} choosing $f(R)=\cosh R$ and $m(R)=r_{g} +a_{0} \sinh ^{3} R$ one obtains

\begin{equation} \label{GrindEQ__62_} 
\begin{array}{c} {r(R,t)=\frac{r_{g} +a_{0} \sinh ^{3} R}{\sinh ^{2} R} \sinh ^{2} \frac{\alpha }{2} ;} \\ {t-t_{0} (R)=\frac{r_{g} +a_{0} \sinh ^{3} R}{2\sinh ^{3} R} (\sinh \alpha -\alpha ).} \end{array} 
\end{equation}

Thus using the mass function method we have obtained new exact solutions \eqref{GrindEQ__58_}, \eqref{GrindEQ__61_}, \eqref{GrindEQ__62_} for the black hole on the dust cosmological background for the flat, closed and open cases respectively.

\subsection*{Conclusions}

The use of mass function method significantly simplifies the problem of obtaining the exact solutions of the Einstein equations. 

The mass function method was demonstrated in application to known exact solutions for the empty space. New exact solutions such as the solution in Gaussian coordinate system (\ref{GrindEQ__11_}) and Kruskal-Szekeres generalized solution (\ref{GrindEQ__20_}) were obtained. 

It was shown that the mass function method makes it much simpler to obtain the Tolman-Bondi solutions. Exact Friedmann-like solutions were obtained for the case of nonzero pressure (\ref{GrindEQ__391_} -- \ref{GrindEQ__393_}). Besides the solutions for Schwarzschild-like black holes were considered in terms of mass function. It was shown that such black holes necessarily contain non-baryonic matter. New exact solutions for the black hole embedded into the dust universe were obtained (\ref{GrindEQ__59_} -- \ref{GrindEQ__62_}).

\end{document}